\documentclass[english,twocolumn,aps,prl,showpacs]{revtex4-1}

\usepackage[T1]{fontenc}
\usepackage[latin1]{inputenc}
\usepackage{graphicx}
\usepackage{amssymb}
\usepackage{epsfig}

\makeatletter

\input{epsf}

\makeatother

\usepackage{babel}
\makeatother

\begin{document}

\title{Crossover from 2D to 3D in a weakly interacting Fermi gas}

\author{P. Dyke,$^1$ E. D. Kuhnle,$^1$ S. Whitlock,$^{1,2}$ H. Hu,$^1$ M. Mark,$^1$ S. Hoinka,$^1$ M. Lingham,$^1$ P. Hannaford,$^1$ and C. J. Vale$^1$}
\affiliation{$^1$ARC Centre of Excellence for Quantum-Atom Optics, Centre for Atom Optics and Ultrafast Spectroscopy, \\
Swinburne University of Technology, Melbourne 3122, Australia\\  $^2$Physikalisches Institut, Universit\"{a}t Heidelberg, Philosophenweg 12, 69120 Heidelberg, Germany }

\date{\today}

\begin{abstract}
We have studied the transition from two to three dimensions in a low temperature weakly interacting $^6$Li  Fermi gas.  Below a critical atom number, $N_{2D}$, only the lowest transverse vibrational state of a highly anisotropic oblate trapping potential is occupied and the gas is two-dimensional.  Above $N_{2D}$ the Fermi gas enters the quasi-2D regime where shell structure associated with the filling of individual transverse oscillator states is apparent.  This dimensional crossover is demonstrated through measurements of the cloud size and aspect ratio versus atom number.
\end{abstract}

\pacs{03.75.Ss, 03.75.Hh, 05.30.Fk, 67.85.Lm}

\maketitle

Dimensional crossovers provide a bridge between seemingly disparate behaviors in physical systems.  For example, Bose-Einstein condensation (BEC) and superfluidity generally coexist in three dimensions (3D) but, in 2D, Berezinskii-Kosterlitz-Thouless (BKT) superfluidity can occur without BEC \cite{berezinskii71,kosterlitz73}.  Ultracold atomic gases offer unprecedented opportunities to explore such crossovers where the dimensionality and interactions can be precisely controlled \cite{bloch08}.  The evolution from a BEC to a BCS (Bardeen-Cooper-Schrieffer) superfluid at a Feshbach resonance is now a central topic in 3D Fermi gas research \cite{jochim03, greiner03, bourdel04, zwierlein05, partridge05,kinast05,giorgini08}.  Restricting these gases to 2D dramatically alters pairing and superfluidity \cite{randeria90,petrov01,petrov03,bothelo06,salasnich07,zhang08a,iskin09,zhang08b,tempere09} and may offer insight into itinerant ferromagnetism \cite{jo,conduit} and Fulde-Ferrell-Larkin-Ovchinikov superfluidity \cite{fulde,larkin,liao}, also seen in quasi-2D heavy fermion superconductors \cite{kumagai}.  In the crossover region, the evolution from a BCS to BKT superfluid can also be investigated.

Dimensional crossovers are characterized by access to new degrees of freedom.  As new quantum states become accessible, their discrete energies can be immediately apparent.  In quasi-2D Fermi gases, where the transverse confinement energy is on the scale of the Fermi energy, occupation of new transverse states gives rise to shell structure, leading to steps in the density profile, chemical potential and specific heat with increasing system size as a consequence of Pauli exclusion \cite{schneider98,vignolo03,mueller04}.

In this letter we study the crossover from 2D to 3D in a two-component optically trapped $^6$Li Fermi gas.  The scaling of the cloud width with atom number in both the tight and weakly confined directions changes dramatically through the crossover.  We also see clear indications of shell structure associated with the filling of discrete energy levels, a feature which is most pronounced in the aspect ratio of the cloud.  The experimentally observed structure agrees well with theoretical predictions for a weakly interacting Fermi gas.

Recent experiments have investigated 2D phenomena in Fermi gases prepared in one-dimensional optical lattices \cite{modugno03,gunter05,du09,martiyanov10,frohlich10}.  Lattice-based experiments produce multiple 2D clouds that can be imaged simultaneously offering good measurement signal-to-noise.  However, the cloud size in the tightly confined direction is typically sub-micron, well below the resolution of nearly all imaging systems, so the properties of individual clouds cannot be easily measured.  We overcome this limitation by creating single 2D clouds, that can be expanded for imaging.

Achieving the 2D regime in a Fermi gas requires that the Fermi energy, $E_F$, and temperature, $T$, are sufficiently low that excitations in one dimension ($z$) are forbidden.  This can be understood by considering an ideal Fermi gas confined in an harmonic trapping potential
\begin{equation}
V(x,y,z) = \frac{1}{2}m(\omega_r^2 x^2 + \omega_r^2 y^2 + \omega_z^2 z^2),
\end{equation}
where $m$ is the mass of the atoms and $\omega_{r,z}$ are the trapping frequencies in the radial ($r$) and transverse ($z$) directions, respectively.  For simplicity, we consider the radially symmetric case ($\omega_x = \omega_y = \omega_r$) and an oblate geometry where $\omega_z \gg \omega_r$.  In 2D the Fermi energy is given by $E_{F,2D}= \sqrt{2N} \hbar \omega_r$ where $N$ is the number of atoms in each spin state.  In 3D the Fermi energy is $E_{F,3D}= (6N)^{1/3} \hbar \bar{\omega}$ where $\bar{\omega} = (\omega_x \omega_y \omega_z)^{1/3}$.

At zero temperature, the Fermi radii are set by the Fermi energy.  In 3D this gives $R_{F,r_i} = (48N)^{1/6} \sqrt{\hbar \bar{\omega}/(m \omega_{r_i}^2)}$, where $r_i = x,y,z$.  In contrast, when the gas is two-dimensional ($E_F \ll \hbar \omega_z $), the width in the tightly confined direction is set by the size of the harmonic oscillator ground state $a_z = \sqrt{\hbar/(m \omega_z)}$.  In the radial directions, however, the Fermi radius is
\begin{equation}
R_{F,r} = (8N)^{1/4} \sqrt{\frac{\hbar}{m\omega_r}}.
\label{eq:fermi_radii_2d}
\end{equation}
This $N^{1/4}$ growth is more rapid than in 3D as two phase-space degrees of freedom ($z, p_z$) are no longer accessible.  As $E_F$ depends on the atom number we define a critical number $N_{2D}$ below which atoms populate only the lowest transverse vibrational state at $T = 0$ and the gas is 2D.  This is equal to the number of single particle states with energy less than the lowest state with one transverse excitation. Labeling harmonic oscillator states by the vibrational quantum numbers $n_r$ and $n_z$ and defining the trap aspect ratio as $\lambda = \omega_z / \omega_r$  we can count the number of ($n_z = 0$) states with energy less than the $(n_r = 0, n_z = 1)$ state.  These states satisfy $n_r < \lambda$ so, including degeneracies, $N_{2D}$ is given by
\begin{equation}
N_{2D} = \sum_{n_r = 0}^{\lambda-1} (n_r + 1) = \frac{\lambda}{2} (\lambda + 1).
\end{equation}
In the experiments that follow, we work in a trap with $\lambda \approx 60$ corresponding to $N_{2D} \approx 1800$.  

When only a few transverse states are occupied the gas is quasi-2D and the number $N_{n}$ at which the $n^{\mathrm{th}}$ transverse state begins to fill is given by 
\begin{equation}
N_n = \frac{\lambda}{4} n(n+1)\left[ \frac{\lambda}{3} (2n+1) + 1 \right].
\end{equation}
At these points we expect steps in the scaling of the cloud size with $N$, giving rise to shell structure.

We quantify the cloud size using root mean square (rms) radii $\sigma_{r_i} = \sqrt{\langle r_i^2 \rangle}$ which provide a model independent width measure applicable at both zero and finite temperatures.  To calculate $\sigma_{r_i}$ theoretically, we first find the 3D density profile $n(x,y,z)$, integrate this over two dimensions to obtain a line profile $n(r_i)$ and then evaluate the second moment $\langle r_i^2 \rangle = \int n(r_i) r_i^2 dr_i / \int n(r_i) dr_i \equiv \sigma_{r_i}^2$.  For an ideal Fermi gas, $n(x,y,z)$ is found by summing the squared wavefunctions of the individual oscillator states.  However, as our experiments are performed at a magnetic field of 992 G where the $s$-wave scattering length is $a_s = -4300 a_0$ ($a_0$ is the Bohr radius), we have also calculated $n(x,y,z)$ by numerically solving the 3D Hartree-Fock mean-field equations for our oblate trapping potential. In the weakly confined radial directions, we use a local density approximation assuming a slowly varying density profile as a function of $r$.  At each $r$, we then solve the Schr\"{o}dinger equation for the transverse direction to obtain the full density profile $n(x,y,z)$.

To create a 2D Fermi gas, we begin with a cloud of approximately $N = 10^5$ $^6$Li atoms in each of the lowest two spin states $|F = 1/2,m_F = \pm 1/2\rangle$ in a far detuned optical dipole trap.  The cloud is evaporatively cooled to a temperature $T \approx 0.1 \, T_F$ at a magnetic field 834 G \cite{fuchs07}, at the centre of the Feshbach resonance, where elastic collisions are unitarity-limited.   To vary the final atom number we continue the evaporation by further lowering the dipole trap power so that atoms are spilled while maintaining the cloud at the lowest possible temperature.  

Next we adiabatically ramp on the 2D optical trap, formed by a cylindrically focussed Gaussian beam propagating along the $y$-direction with $1/e^2$ waists of $w_z = 8 \,\mu$m and $w_x = 400 \,\mu$m in 200 ms.  Similar configurations have been used to study 2D Bose gases \cite{gorlitz01,clade09}.  Once this trap is fully on, the first beam used for evaporation is ramped down in 200 ms leaving the atoms in the 2D trap.  Confinement in the $y$-direction  is achieved by the short Rayleigh length associated with the 8 $\mu$m waist.  Additional confinement in the $x$ and $y$-directions is provided by the curvature of the magnetic field.  The 2D trapping frequencies are $\omega_z/2\pi= 2800$ Hz and $\omega_r/2\pi= 47$~Hz ($\omega_x \sim \omega_y \equiv \omega_r$) giving an aspect ratio of approximately 60.  Finally we adiabatically ramp the magnetic field to 992 G where the cloud is imaged.  Precise temperatures in the 2D trap are difficult to ascertain due to the lack of an analytic model for interacting quasi-2D gases.  However, we expect the temperature to be below $0.1 \, T_F$ due to the adiabatic field sweep and deep evaporation used to prepare the low atom number clouds. 

To demonstrate the crossover from 2D to 3D we measure the cloud radii in the tight and weakly confined directions as a function of $N$.  In the $z$-direction the cloud width in trap ($a_z \approx 770$ nm) is much smaller than the resolution of our imaging system, so a short time of flight ($500 \,\mu$s) is used before imaging.  This time is long compared to the inverse trapping frequency in the $z$-direction ($1/\omega_z = 57 \,\mu$s) but short compared to $1/\omega_y$ (3.4 ms).  The cloud distribution in the radial direction will therefore be equivalent to the in-trap distribution.

    \begin{figure}[!t]
        \centering
        \includegraphics[clip,width=0.48\textwidth]{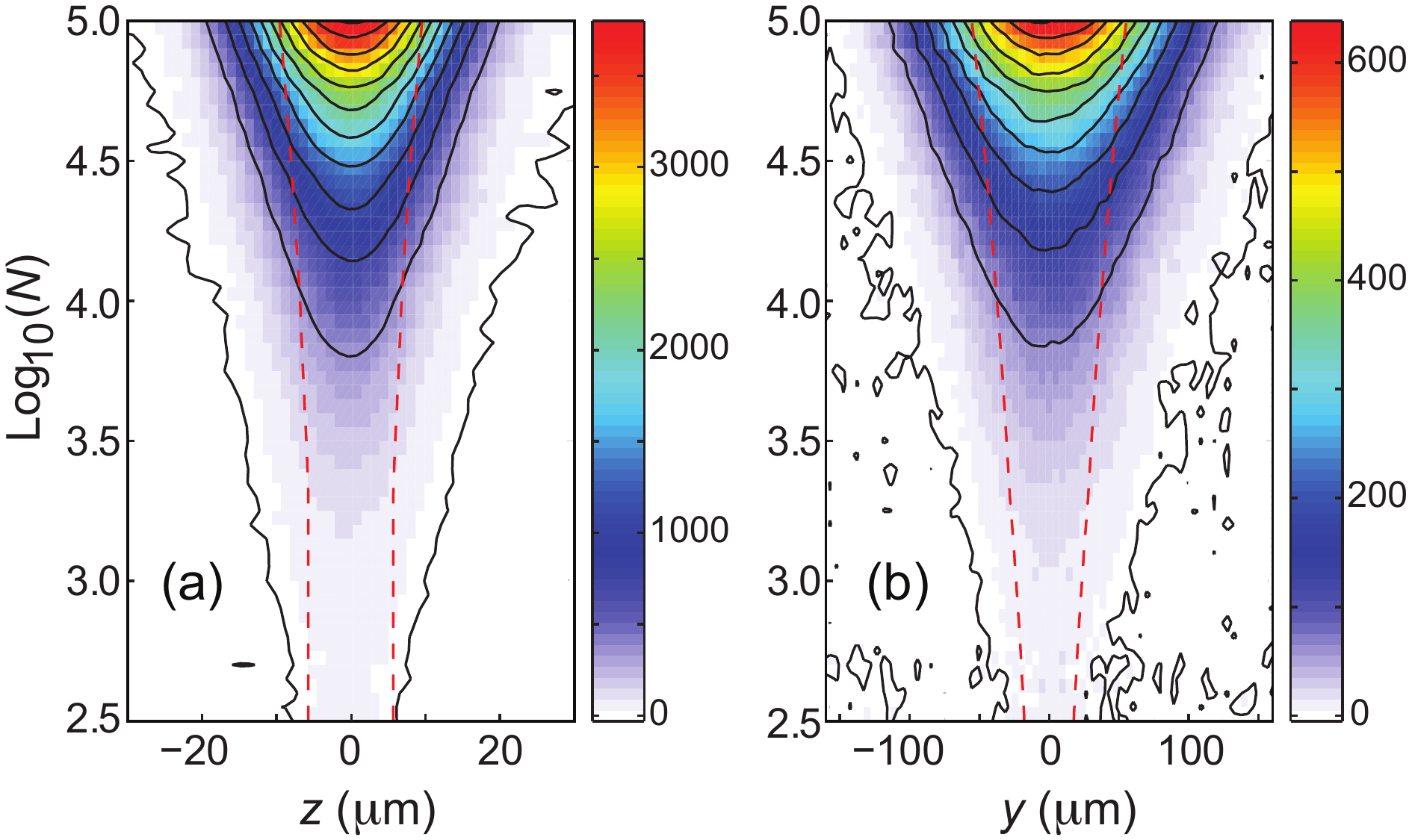}
        \caption{One-dimensional density profiles for clouds at 992~G as a function of atom number in the tight (a) and weakly confined (b) directions after 500 $\mu$s expansion.  Solid black lines are contours of equal density and the dashed red lines show the predicted rms cloud width for an ideal Fermi gas.  The color bar indicates line density in units of atoms/$\mu$m.}
        \label{fig:fig1}
    \end{figure}

The imaging beam propagates roughly along the (radial) $x$-direction, so we obtain two-dimensional density profiles $n(z,y)$.  Absorption images are processed using a fringe removal algorithm \cite{ockeloen10} to optimize image quality.  From these images we generate line profiles $n(z)$ and $n(y)$ by integrating over the other dimension.  Figure 1 shows the measured line-densities as a function of atom number for gases at 992 G ($a_s = -4300 a_0$) in both the tight and weakly confined directions.  The plots consist of approximately 50 profiles; each profile is the average of several images binned according to $N$.  The dashed red lines show scaled predictions for the zero temperature rms radius for an ideal Fermi gas and the solid lines are contours of equal density.

In order to compare these data with theory, we evaluate the rms widths from the profiles in Fig. 1.  Figure 2 (main panel) shows the rms cloud width in the $y$-direction ($\sigma_y$) versus atom number.  Data points are the experimental measurements and the solid and dashed lines are the predictions for a weakly interacting and ideal Fermi gas, respectively.  Each theory curve has been scaled by 1.19 (1.11) for the weakly interacting (ideal) gas to lie on top of the experimental data at low $N$, where interactions are least significant the models are most accurate.  These scalings account for factors such as finite imaging resolution, nonzero cloud temperature and a slight ellipticity in the radial confinement ($\omega_x / \omega_y \approx 1.1$), but do not affect the power-law dependence of the width.  The data closely follow the predicted growth rate for a weakly interacting gas over the full range of atom numbers with an elbow around $N = 2000 \, (\approx N_{2D})$ corresponding to the transition between 2D and quasi-2D regimes.  The data deviate below the ideal gas prediction at high $N$ where interactions become more important.   As the atom number and density increases, the attractive interactions become more significant and slow the growth rate of the cloud.  For the lowest numbers ($N \approx 800$) the interaction parameter $1/(k_F a_s) = -2.3$, where $k_F$ is the Fermi wave-vector.  For $N \approx 10^5$ the gas is approaching the strongly interacting regime, $1/(k_F a_s) = -1.0$, and our weakly interacting theory will begin to break down.  Below $N_{2D} = 1800$, a least squares fit to the experimental data shows that the width grows as $N^{0.28 \pm 0.05}$, in agreement with the $N^{0.25}$ scaling predicted by Eq. (2).  Above $N_{2D}$, the fitted dependence is $N^{0.151 \pm 0.004}$, in reasonable agreement with the interacting gas prediction of $N^{0.146}$ but well below the ideal gas prediction of $N^{0.174}$. 

    \begin{figure}[!t]
        \centering
        \includegraphics[clip,width=0.46\textwidth]{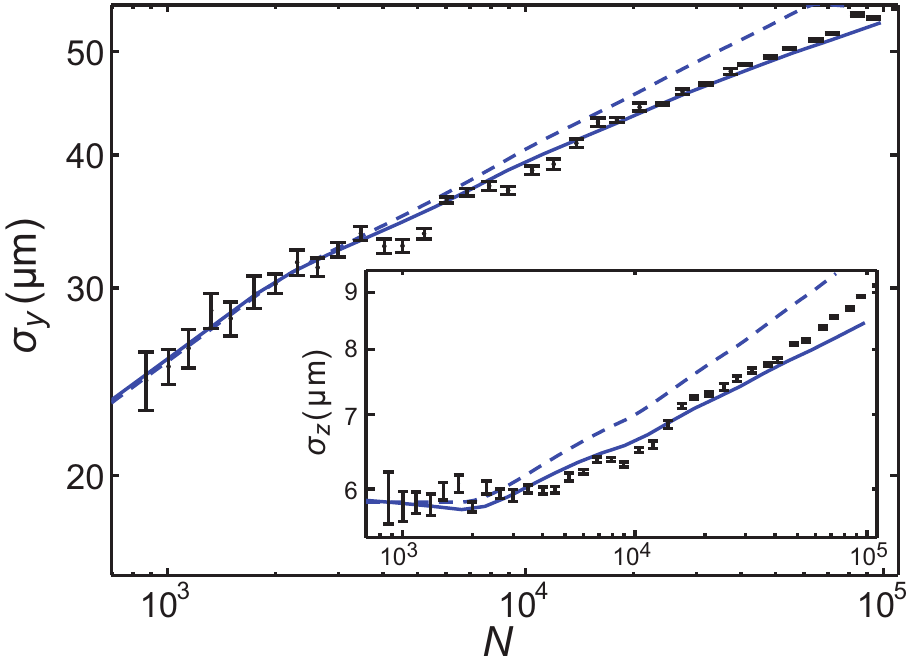}
        \caption{Measured rms cloud widths (data points) and theoretical predictions for a weakly interacting  (solid lines) and ideal (dashed lines) Fermi gas.  Radial cloud width is plotted in the main panel and the inset shows the transverse cloud width after 500 $\mu$s expansion.  Theoretical curves are scaled as described in the text.}
        \label{fig:fig2}
    \end{figure}

The inset of Fig. 2 shows the cloud width in the $z$-direction.  The theoretical widths here have been scaled by $b \sqrt{1+\omega_z^2 t^2}$ to account for the $t = 500 \,\mu$s expansion.  The value of $b$ which provides the best agreement with the low $N$ data is 1.24 (1.20) for the weakly interacting (ideal gas) theory.  These values are slightly larger than those used in the radial direction suggesting the expansion may not be simply ballistic.  Below $N_{2D}$ the transverse width is roughly constant scaling as $N^{0.07 \pm 0.04}$ in reasonable agreement with the predictions of $N^{-0.01}$ for an interacting gas and $N^{0}$ for an ideal gas.  Finite temperature may lead to occupation of the ($n_z = 1$) state for $N < N_{2D}$ which would increase the measured exponent.  For $N > N_{2D}$ the width scales as $N^{0.124 \pm 0.004}$, compared to $N^{0.100}$ and $N^{0.133}$ for the interacting and ideal gases, respectively.  These exponents measured in the $z$-direction may be influenced by the cloud expansion which could have a non-trivial dependence on $N$ \cite{menotti02,hu04} as $1/(k_F a_s)$ varies by more than a factor of 2 for the range of atom numbers considered here.  We note that the calculated exponent for the non-interacting gas is below the true 3D value of 1/6 as the data lie in the quasi-2D regime where shell structure influences the widths. 

To further investigate the shell structure we now focus on the cloud aspect ratio, $\kappa =  \sigma_z / \sigma_y$.  Shell structure will be more prominent in measurements of $\kappa$ than in the individual widths as the filling of new transverse shells begins in states with low radial quantum numbers; hence an increase in the transverse cloud size will correlate with a decrease in the growth rate of the radial size. In the 3D limit, the cloud aspect ratio would be constant but in 2D and quasi-2D $\kappa$ will show a strong dependence on the atom number.  Additionally, as $\kappa$ is given by the ratio of two measured quantities, certain experimental systematics (e.g. finite imaging resolution and shot to shot temperature variations) will be reduced.  

In Fig. 3 (main panel) we plot the aspect ratio of the cloud along with theoretical predictions for the weakly interacting (solid line) and ideal (dashed line) Fermi gases.  The agreement between theory and experiment is very good and the large change in aspect ratio with $N$ clearly demonstrates departure from 3D behavior.  Only at high atom numbers does the aspect ratio level off indicating broad coverage of the 2D and quasi-2D regimes.  The arrows indicate the calculated atom numbers at which new transverse states become accessible, Eq. (4).  The interacting and ideal gas predictions are very similar, emphasizing the robustness of aspect ratio measurements for identifying the dimensional crossover.  

    \begin{figure}[!t]
        \centering
        \includegraphics[clip,width=0.46\textwidth]{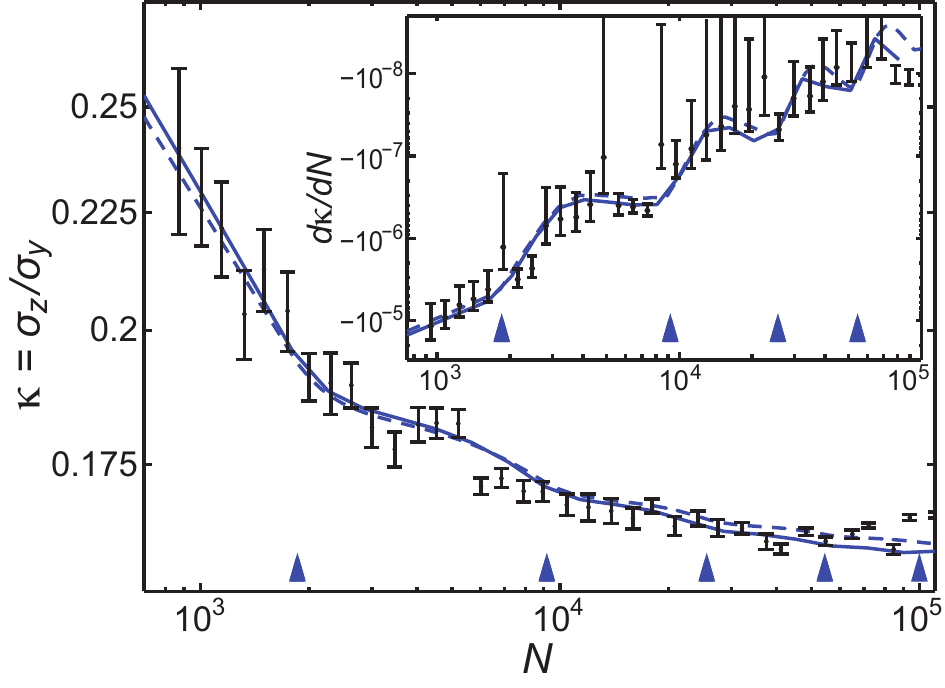}
        \caption{Cloud aspect ratio in the 2D-3D crossover (points) and theoretical predictions for a weakly interacting (solid line) and ideal (dashed line) Fermi gas.  Arrows indicate predicted atom numbers at which new transverse states become accessible. Inset: Gradient of the experimental and theoretical aspect ratios showing the signature of shell-filling.}
        \label{fig:fig3}
    \end{figure}

At low $N$, the aspect ratio decays steadily with increasing atom number, before a step occurs at $N_{2D}$ corresponding to occupation of the first transverse excited state.  The inset of Fig. 3 shows the gradient $d\kappa/dN$ of the theoretical and experimental aspect ratios which show the signatures of shell structure more clearly.  The experimental gradients were evaluated numerically and smoothed with a five-point moving average.  Both the measured aspect ratio and gradient closely follow the theoretical predictions, with indications of shell structure present in the experimental data for $N \lesssim 10,000$ corresponding to occupation of the ground and first transverse excited states.  The location of the first two arrows in the plots agrees well with the position of the steps.   For larger $N$ the data lie close to the theoretical line but the shell structure is unresolved. 

This work provides the first quantitative study of the transition from 2D to quasi-2D and 3D in a weakly interacting Fermi gas.  At low atom numbers, shell structure, associated with the filling of individual transverse oscillator states, becomes apparent.  Our data were obtained away from any confinement induced scattering resonances ($a_s < a_z$); however, these could easily be accessed closer to the Feshbach resonance \cite{petrov01,frohlich10,haller10}.  This work opens the way to investigations of the phase diagram of 2D and quasi-2D Fermi gases as a function of the scattering length and temperature and could help elucidate the evolution from BKT to BCS superfluidity through the 2D-3D crossover.

This work is supported by the Australian Research Council Centre of Excellence for Quantum-Atom Optics.

\end{document}